\documentclass[prb,twocolumn]{revtex4}
\usepackage{graphicx}
\usepackage{amsmath,amssymb}
\begin{document}
\title{Gradient corrections to the exchange-correlation free energy}
\author{Travis Sjostrom and J\'er\^ome Daligault}
\affiliation{Theoretical Division, Los Alamos National Laboratory,
Los Alamos, New Mexico 87545}
\date{\today}
\begin{abstract}
  We develop the first order gradient correction to the exchange-correlation free energy of the homogeneous electron gas for use in finite temperature density functional calculations. Based on this we propose and implement a simple temperature dependent extension for functionals beyond the local density approximation. These finite temperature functionals show improvement over zero temperature functionals as compared to path integral Monte Carlo calculations for deuterium and perform without computational cost increase compared to zero temperature functionals and so should be used for finite temperature calculations.
\end{abstract}
\maketitle
\section{Introduction}

Understanding of matter in extreme conditions represents a significant and current challenge of high energy density physics. Some particular systems of interest include dense astrophysical plasmas as exist in the interiors of giant planets, as well as warm dense matter, which is increasingly studied in high energy density laboratory experiments. In these conditions of elevated temperature and density relative to the ambient condensed matter state, ions can be strongly coupled and electrons are at least partially degenerate. These conditions have proven difficult to describe theoretically and necessitate numeric simulations. One key approach is molecular dynamics simulations via density functional theory (DFT). In DFT the exchange-correlation free energy is a key input approximated by a density functional that is in general temperature dependent.  However, while the DFT approach is increasingly used to study higher temperature systems, zero temperature exchange-correlation functionals are most commonly employed as opposed to explicitly temperature dependent functionals.

Recently fits were provided for the finite temperature exchange-correlation local density approximation (LDA) \cite{ftlda} which is the simplest type of density functional. At zero temperature the LDA has seen significant improvements made upon it over the past 40 years. In the first step beyond LDA density gradient expansions were examined, then generalized gradient approximations were developed, and later even more complex, orbital-dependent functionals were considered \cite{Perdew}. While a similar effort has not been seen at finite temperature, Geldart and co-workers \cite{Geldart} did derive the gradient expansion for the exchange only contribution. In this paper we examine the gradient expansion for the full exchange-correlation functional and based on that provide a simple finite temperature extension for generalized gradient functionals. In addition we perform self-consistent calculations to determine the overall importance of temperature dependence in exchange-correlation functionals.

\section{Gradient corrections to the exchange-correlation free energy}

\subsection{Development of the gradient expansion}
In order to determine the gradient expansion we consider the relation of density functional theory to dielectric theory for the uniform electron gas. Following Kohn and Sham we first write the gradient expansion of the exchange-correlation free energy as
\begin{align}
  F_{\mathrm xc}[n] = &\int  d{\mathbf r} f_{\mathrm xc}(n) n(\mathbf r) \nonumber \\
&+ \frac{1}{2} \int  d{\mathbf r} g_{\mathrm xc}^{(2)}(n) \left| \nabla n(\mathbf r)\right|^2 + \dots \;.
\end{align}
The first term of the RHS on its own is the local density approximation, with $f_{\mathrm xc}$ the exchange-correlation free energy per electron in the uniform electron gas. The coefficient of the gradient correction term, $g_{\mathrm xc}^{(2)}$, is the piece determined in this work, and it is related to the static local field correction $G(k)$ of the homogeneous electron gas by\cite{Gupta,Niklasson,KohnSham,HohenbergKohn}
\begin{align}
  g_{\mathrm xc}^{(2)}(n) &= \frac{1}{2} \left( 
\frac{\partial^2 [-v_k G(k)]}{\partial k^2}
\right)_{k \rightarrow 0} \nonumber \\
&= -4 \pi e^2 \delta / k_F^4 \;,
\label{eq:gxc2}
\end{align}
where $v_k= 4 \pi e^2 / k^2$ is the Coulomb potential, $k_F=(3\pi^2n)^{1/3}$, and in the second line we consider the small $k$ expansion of $G(k) = \gamma (k/k_F)^2 + \delta (k/k_F)^4+\dots$ Here the dependence of $G$, $\gamma$  and $\delta$ on the electron density, $n$, and temperature, $T$, is suppressed for convenience.

It is known that $G$ may be well represented for small and large $k$, though not for intermediate values, by the function \cite{VS,DAC}
\begin{align}
  G(q)=A\left( 1-e^{B q^2}\right)
\end{align}
with $q=k/k_F$. Then for small $q$
\begin{align}
  G(q)=ABq^2-\frac{1}{2}AB^2q^4+\dots \;,
\end{align}
and so,
\begin{align}
  \gamma=AB \quad\quad \delta=-AB^2/2\;.
\end{align}
Next from the compressibility sum rule we have
\begin{align}
  \gamma=-\frac{k_F^2}{4\pi e}\frac{\partial^2 (n f_{xc})}{\partial n^2}\;,
\end{align}
which we may evaluate by the recent analytic fits \cite{ftlda} to the the quantum Monte Carlo (QMC) data \cite{Brown}. This leaves us needing still $A$ or $B$ independently to determine $\delta$ and hence $g_{xc}^{(2)}$. This is completed then by the relation for the large $q$ limit of $G(q)$ to the pair distribution function \cite{TanakaIchimaru},  $g(r$),
\begin{align}
  A=G(q\rightarrow \infty)=1-g(0)\;.
\label{eq:g0}
\end{align}

\begin{figure}
  \includegraphics[width=1.0\columnwidth]{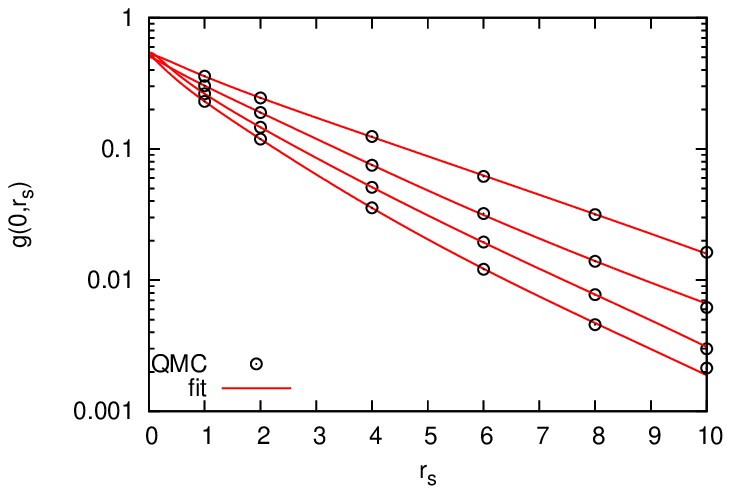}
   \includegraphics[width=1.0\columnwidth]{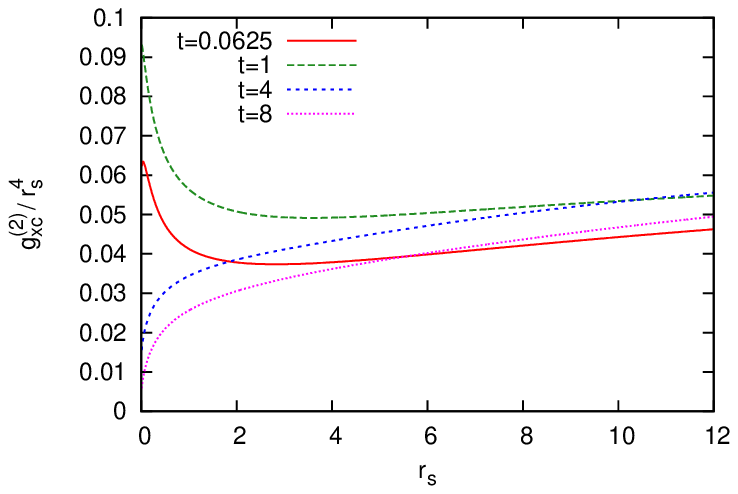}
  \caption{Top: QMC data points for $g(0)$ and our fit as a function of $r_s$ for a given $t=k_BT/E_F$, with the the curves being $t=8,4,0.0625,1$ from top to bottom. 
Bottom: Coefficient of the gradient expansion $g_{xc}^{(2)}$, for given $t$ as derived from QMC fits for $f_{xc}$ \cite{ftlda} and $g(0)$ (top).}
\label{fig:g0}
\end{figure}

In order to determine the gradient coefficient, we now need only $g(0)$ further. This we obtain from the recent restricted path integral Monte Carlo results of Brown et al. \cite{Brown} in which they provide $g(r)$ for the unpolarized system. However their grid does not include $g(0)$ so we have extrapolated their $g(r)$, according to the cusp condition \cite{Kimball}, using the form $g(r)= a + ar + br^2$ for small $r$ to obtain $g(0)$ at each density and temperature point given in the QMC results. Next we fit $g(0)$ as a function of $r_s$ for each temperature $t=k_BT/E_F=\{0.0625, 0.125, 0.25, 0.5, 1.0, 2.0, 4.0, 8.0\}$, according to the following equation \cite{Spinketal}
\begin{align}
  g(0)= \frac{1}{2} \frac{1+a \sqrt{r_s}+b r_s}{1 + c r_s + d r_s^3}
\end{align}
The fit results are plotted for selected $t$ in the upper panel of Fig. \ref{fig:g0} along with the QMC data. 
Then using Eqs. \ref{eq:gxc2}-\ref{eq:g0} along with the fits for $g(0)$ we find $g_{xc}^{(2)}$ as a function of $r_s$ for the given $t$ values. The results are plotted for $t=0.0625,1,4,8$ in the lower panel of Fig. \ref{fig:g0}. It is worth noting that the $t=0.0625$ result for $g(0)$ is nearly identical to the $t=0$ QMC result from Spink et al. \cite{Spinketal}.

\subsection{Analysis of the temperature dependence}

\begin{figure*}
  \includegraphics[width=0.68\columnwidth]{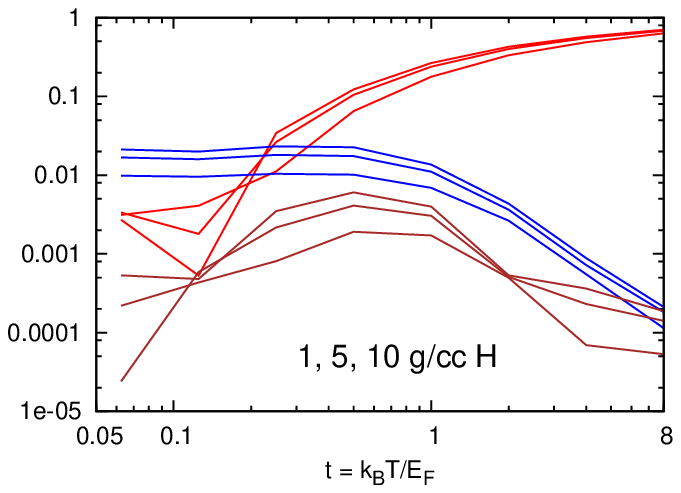}
  \includegraphics[width=0.68\columnwidth]{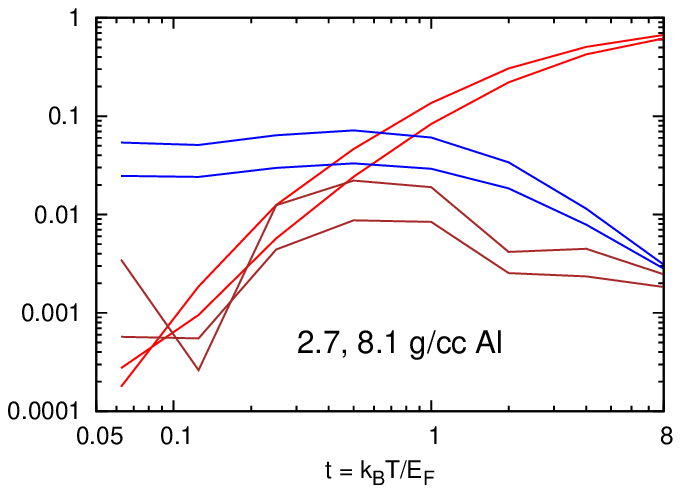}
  \includegraphics[width=0.68\columnwidth]{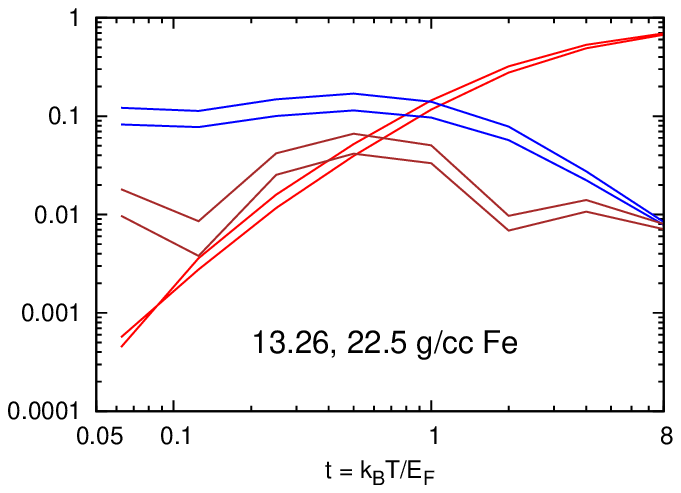}
  \caption{Relative effects of the terms of the gradient expansion for the exchange-correlation free energy. The temperature dependence of the local density term becomes dominant at high temperatures (red curves), and the gradient term is most important at lower temperatures (blue curve), while the temperature dependence of the gradient term is always negligible (brown curves). Different curves of the same color represent the different densities.}
\label{fig:relxc}
\end{figure*}

To examine the effects of the temperature dependent gradient coefficient, $g_{xc}^{(2)}$, we calculate its relative contributions on various systems at different temperatures and densities. Specifically we solve all electron hydrogen, aluminum, and iron systems  at each $t$ for which we have fit $g_{xc}^{(2)}$ and at selected densities from ambient to several times ambient compression. We first solve the system in an average atom model \cite{Feynmann} using an orbital-free functional for the non-interacting contribution, namely the Thomas-Fermi plus von Weisz\"acker approximation, and a zero temperature LDA for the exchange-correlation energy. This gives us a realistic density. Then using this density we evaluate the different exchange-correlation free energy contributions to determine their relative effects, with the results shown in Fig. \ref{fig:relxc}. The relative effect of the temperature dependence of the local density term is shown in the red curves given by
\begin{align}
  \frac{\int d\mathbf{r}\; n \left( f_{xc}-\epsilon_{xc} \right) }{\int d\mathbf{r}\; n \epsilon_{xc}} \;,
\end{align}
where $\epsilon_{xc}=f_{xc}(t=0)$ is the zero temperature exchange-correlation energy.
While the total relative contribution due to the gradient term as well as that portion due to its finite temperature contribution are given by
\begin{gather}
  \frac{\int{d\mathbf{r}\; \left| \nabla n \right|^2 g_{xc}^{(2)}/2}}{\int d\mathbf{r}\; n f_{xc}} \;,\\
\frac{\int{d\mathbf{r}\; \left| \nabla n \right|^2 \left[ g_{xc}^{(2)}(t) - g_{xc}^{(2)}(t=0)\right]/2}}{\int d\mathbf{r}\; n f_{xc}}\;,
\end{gather}
and shown in the blue and brown curves respectively. Theses results show that at high temperature, dependence of the local density term provides the dominant correction, however at low temperature the gradient correction is most important. By comparison the temperature dependent correction coming from the gradient term is always negligible.

\subsection{Beyond the gradient expansion}

In considering an improved gradient corrected functional for the exchange-correlation free energy we reiterate that the previous analysis shows that temperature dependence in the gradient term is in fact negligible. That is using a zero temperature gradient correction at finite temperature is in fact a good approximation. However this applies only to the gradient term, the local density term does show a significant temperature dependence and a proper finite temperature functional should be used in that case. It is also clear that gradient corrections are important at low temperature and it is well known from zero temperature development that generalized gradient approximations (GGAs), such as PBE \cite{PBE}, perform significantly better than gradient expansions. Therefore we propose a temperature dependent GGA as follows
\begin{align}
  F_{xc}^{GGA}[n] = E_{xc}^{GGA}[n]-E_{xc}^{LDA}[n]+F_{xc}^{LDA}[n]\;.
  \label{eq:fxcgga}
\end{align}
Here the zero temperature GGA term, $E_{xc}^{GGA}[n]$ includes the local density contribution. Then the zero temperature local density contribution is removed and replaced with the finite temperature version. Thus capturing all significant temperature and density gradient dependence in the exchange-correlation free energy. 

\subsection{Self-consistent results}
We have implemented the finite temperature exchange-correlation free energy in both the local density approximation as given in Ref. \onlinecite{ftlda}, as well as in finite temperature modification of the zero temperature PBE functional according to Eq. \ref{eq:fxcgga} in the plane wave density functional theory code Quantum-Espresso \cite{qespresso} as well as in our orbital-free code. We then applied this to cases of warm dense deuterium, for which there exists path integral Monte Carlo (PIMC) results \cite{Huetal} that do not require an approximate input for the exchange-correlation free energy.

\begin{figure}
  \includegraphics[width=1.0\columnwidth]{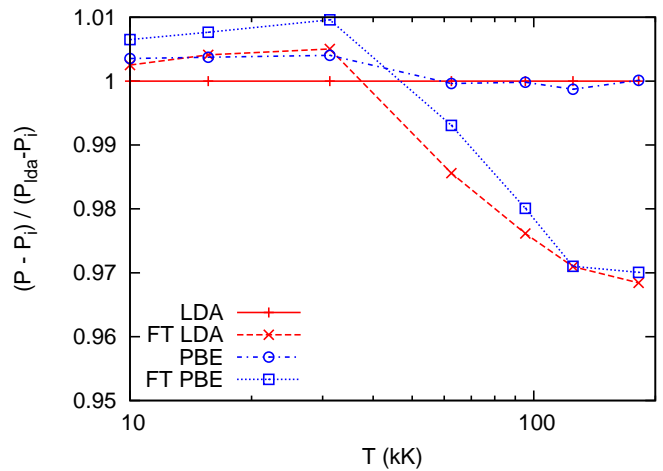}
  \caption{Deuterium pressure, excluding the ion kinetic contribution, at 4.05 g/cc for the LDA and GGA functionals with and without temperature dependence plotted relative to the zero temperature LDA results. Increased significance of the temperature dependent functionals, and decreased effect from the gradient terms for higher temperatures is shown.}
\label{fig:ftldapbe}
\end{figure}

In Fig. \ref{fig:ftldapbe} we plot the resulting pressure relative to the pressure from a zero temperature LDA calculation for deuterium at 4.05 g/cc and up to nearly 200 kK (1000 K = 1 kK). Using the standard zero temperature PBE it is clear that the gradient correction becomes less and less as temperature increases, by contrast both finite temperature functionals show first a small increase in pressure at low temperature, then a more significant decrease in pressure as the temperature is elevated to 200 kK. Again though, the gradient effects diminish with increasing temperature.

\begin{figure}[]
  \includegraphics[width=1.0\columnwidth]{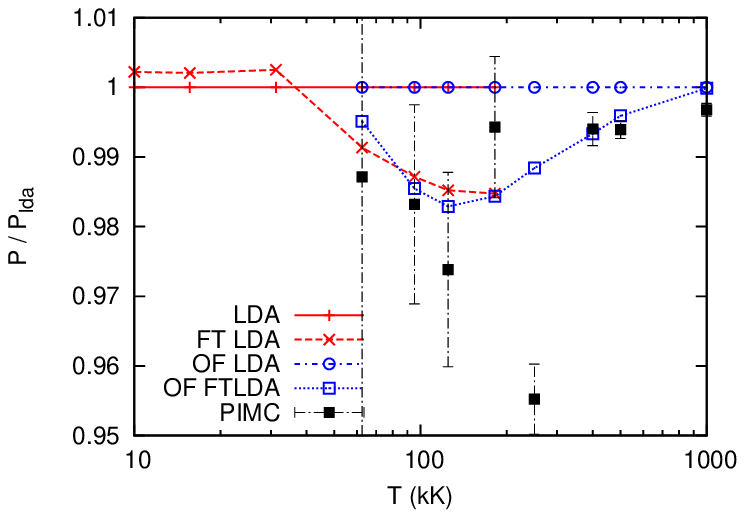}
  \includegraphics[width=1.0\columnwidth]{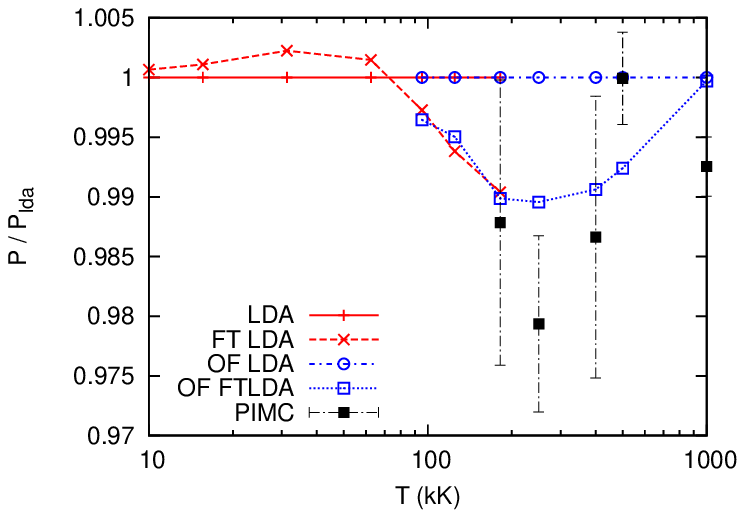}
  \caption{Deuterium pressure at 4.05 g/cc (top) and 10.0 g/cc (bottom) for the LDA functional with and without temperature dependence, as well as PIMC results, relative to the zero temperature LDA results. At both densities the temperature dependent functional agrees better with the PIMC.}
\label{fig:ftlda}
\end{figure}

In order to consider the temperature effect in the warm dense regime, we consider just the LDA functional for deuterium at 4.05 g/cc and at 10.0 g/cc. These results are shown in Fig. \ref{fig:ftlda} along with the PIMC data \cite{Huetal}. In order to extend the calculation from 200 kK up to 1000 kK in temperature we make use of an orbital-free density functional calculation which is seen to be justified as here it overlaps well with the highest temperature Kohn-Sham calculations. The relative pressure results do show better agreement of the temperature dependent functional to that of the PIMC results versus those of the zero temperature functional at both densities. We can see in these cases there is a maximum difference of 1-2\% in the total pressure in the warm dense regime. This effect then diminishes to zero at high temperatures as the total exchange-correlation contribution to the pressure becomes completely negligible compared to the kinetic contributions  of the electrons and ions.

\begin{figure}[]
  \includegraphics[width=1.0\columnwidth]{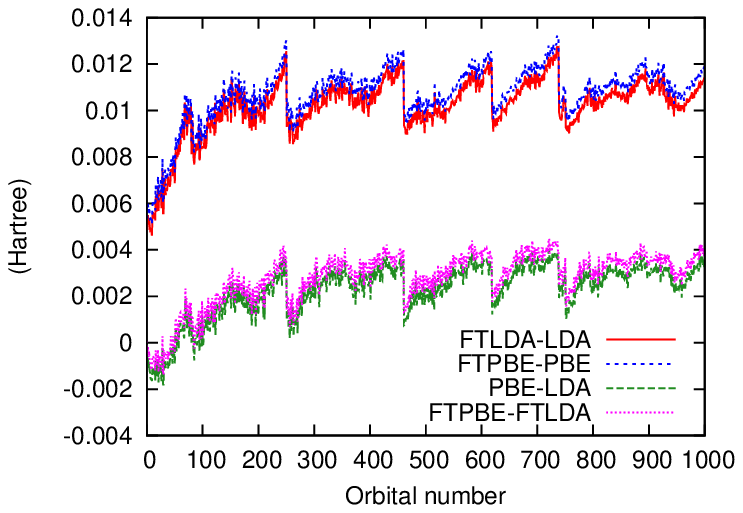}
  \includegraphics[width=1.0\columnwidth]{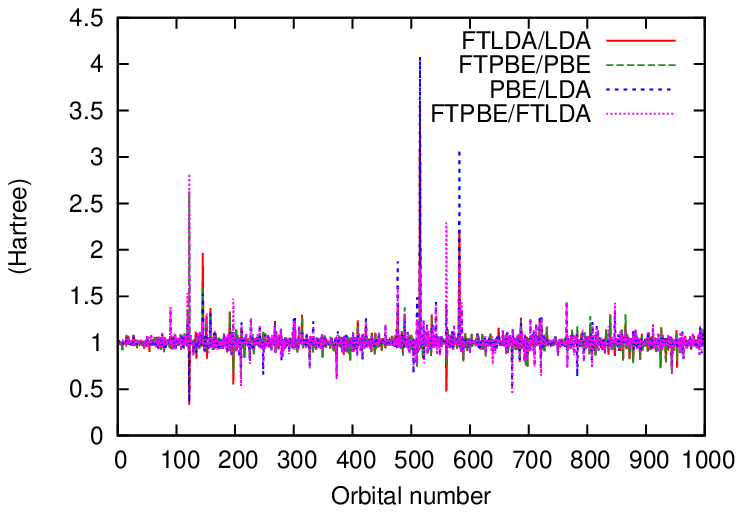}
  \caption{Difference in corresponding eigenvalue energies (upper panel) and the ratio of adjacent eigenvalue differences (lower panel) between different functionals.}
\label{fig:eigs}
\end{figure}

Finally we consider the effect of temperature dependent exchange-correlation on the eigenspectrum of a real system. In the upper panel of Fig. \ref{fig:eigs} we plot for a single random configuration of 128 deuterium atoms at 4.05 g/cc and 15.67 eV the difference in corresponding eigenvalues when using different functionals. The difference between finite temperature and zero temperature functionals here produces about a five times greater difference than the difference between PBE and LDA whether in the finite temperature or zero temperature versions. In the lower panel the ratio of adjacent eigenvalue differences, $(\epsilon_{n+1}-\epsilon_n)$, is taken between the different functionals. The average is clearly seen to always be 1, but there is non-negligible spread seen which is quantified by the standard deviation of 0.12 and 0.09, for the finite temperature to zero temperature functional results of LDA and PBE respectively, and 0.15 and 0.12, for the PBE to LDA results for zero temperature and finite temperature functionals respectively.

\section{Summary}
We derived the finite temperature gradient expansion for the exchange-correlation free energy and then demonstrated that the contribution from the temperature dependence of the gradient term in physical systems is negligible. However the gradient corrections are important at lower temperatures and the finite temperature correction to the local density contribution is important at higher temperatures.
We therefore proposed a temperature dependent GGA and showed that the temperature dependence is more significant than gradient dependence in the warm dense matter regime and that better results are achieved using temperature dependent LDA or GGA, as shown by better agreement with PIMC data for which there is no approximation for the exchange-correlation energy. Finally these finite temperature corrections are easily implemented in any DFT code, through the fit given in Ref. \onlinecite{ftlda} and perform without computational cost increase and so should be used for finite temperature calculations where better accuracy is desired.

\begin{acknowledgments}
  This research has been supported by the DOE Office of Fusion Energy Sciences (FES), and
by the NNSA of the US DOE at Los Alamos National Laboratory under Contract No. DE-AC52-06NA25396. 
\end{acknowledgments}

\end{document}